\documentclass[review]{elsarticle}

\usepackage{lineno,hyperref}
\modulolinenumbers[5]

\usepackage{rotating}
\usepackage{epsfig}          
\usepackage{graphicx}        
\usepackage{amssymb}         
\usepackage{color}           
\usepackage{enumerate}
\usepackage{tabularx}

\journal{Journal of \LaTeX\ Templates}

\begin{document}

\begin{frontmatter}

\title{Solar energetic particle catalogs: assumptions, uncertainties and validity of reports}     


\author[firstaddress]{R. Miteva\corref{mycorrespondingauthor}}

\author[secondaddress]{S.W. Samwel}

\author[thirdaddress]{M.V. Costa-Duarte}


\cortext[mycorrespondingauthor]{Corresponding author.}
\address[firstaddress]{Space Research and Technology Institute, Bulgarian Academy of Sciences, 1113 Sofia, Bulgaria}
\address[secondaddress]{National Research Institute of Astronomy and Geophysics, 11421 Helwan, Cairo, Egypt}
\address[thirdaddress]{Institute of Astronomy, Geophysics and Atmospheric Sciences, University of S\~{a}o Paulo, 05508-090 S\~{a}o Paulo, Brazil}

\begin{abstract}
The aim of this work is to summarize the main underlying assumptions, simplifications and uncertainties while studying solar energetic particles (SEPs). In general, numerous definitions are used for the evaluation of a given SEP parameter and these different methods lead to different outcomes for a given particle event. Several catalogs of SEP events from various instruments are currently available; however, each catalog is specific to the adopted data and analysis. We investigate the differences while comparing several SEP catalogs and outline probable reasons. We focus on SEP statistical studies and quantify the influences of the particle intensity magnitude, solar origin location and projection effects. We found that different definitions and criteria used for these parameters change the values of the correlation coefficients between the SEPs and their solar origin.
\end{abstract}


\end{frontmatter}


\section{Introduction}

Energetic electrons, protons and heavy ions of solar origin are routinely being detected by in situ measurements as particle intensity enhancements. This phenomenon is known as solar energetic particles (SEPs) or solar cosmic rays, see a recent review by \cite{2016LRSP...13....3D}. Energetic particles originate at some solar activity phenomena (e.g., by acceleration processes during solar flares, SFs, and coronal mass ejections, CMEs) because SEPs follow in time these solar eruptions and the SEP time profiles show velocity dispersion characteristics. The energized particles need to escape the acceleration cite (situated in the lower or higher solar corona) and continue to propagate along the heliospheric magnetic field lines. When these field lines sweep over some space-based detector, a particle enhancement is recorded. SEP events are observed by a satellite situated at an isolated point in the heliosphere, either at fixed orbit close to L1 or around the Sun, as the twin STEREO mission \cite{2008SSRv..136....5K}, with rare observations done outside the ecliptic plane (by Ulysses \cite{1992A&AS...92..365S} during the period 1990$-$2009). In this work, we will consider particle data from measurements done close to Earth on a routine basis.

Since 1990s several solar dedicated missions were successfully launched. Among them, SOHO (Solar and Heliospheric Observatory \cite{1995SoPh..162....1D}), ACE and Wind spacecraft continue to provide particle data from their respective instruments. Among these, Energetic and Relativistic Nuclei and Electron (ERNE) instrument \cite{1995SoPh..162..505T} on SOHO and Energetic Particle Acceleration, Composition and Transport (EPACT) instrument \cite{1995SSRv...71..155V} on Wind are considered in the present work. Another prominent example is the series of nearly identical spacecraft located on geostationary orbit, Geostationary Operational Environmental Satellite (GOES), that provide, among others, particle data since late 1970s \cite{2014SW...12..92R}. 

SEP events (usually protons) based on data from the above instruments, are routinely being identified either by observers or via automatic routines. In this work, we consider the following catalogs of proton events: SEPEM (\cite{2015SpWea..13..406C}), GOES-NOAA\footnote{\url{http://umbra.nascom.nasa.gov/SEP/}}, IMP-8 (\cite{2010JGRA..11508101C}), GOES-SSE (\cite{2015SoPh..290..841D}), GOES-SEP (\cite{2016JSWSC...6A..42P}), Wind/EPACT (\cite{2016simi.conf...27M}) and SEPServer (\cite{2013JSWSC...3A..12V}). For the purpose of this study we aim to select several well-known in the SEP community and/or recently compiled proton catalogs. In addition to these, other proton lists also exist\footnote{\url{http://www.wdcb.ru/stp/index.en.html}}, see \cite{SWE:SWE185}, \cite{2009SpWea...7.4008L} and a summary by \cite{2016JSWSC...6A..42P}.

Each of the catalogs reports 100s of individual proton events. Such large amount of data allows for preparing different statistical studies on SEP events and their parent phenomena as well as the quantitative comparisons between the previous and the still ongoing solar cycle (SC), see e.g., \cite{2013AdSpR..52.2102C,2015ICRC....M,2017SunGeo..12.0M}.

The statistical works rely on various catalogs of solar data, usually prepared by different scientific or engineering teams. Numerous assumptions and simplifications are employed during the solar data recording, transfer and processing, in addition to the limitations set by the aging performance of a given instrument. Not accounting for these specifics may lead to erroneous conclusions of the performed scientific analysis. 

We will focus in this study on the period 1996$-$2006 (SC23), since the selected above catalogs have data coverage during this time. Few SEP events are reported during 1996 and none during 2007 and 2008. In this study, we aim to summarize a number of physical, positional, instrumental and observer aspects, which ultimately lead to a different report issued for the same SEP event that affects the overall statistical results. Furthermore, we provide an estimate on the effects of the particle intensity magnitude, solar origin location and projection on the linear correlation coefficients between the particles and their solar origin. In summary, we test the validity of selected well-known SEP results and interpretations that are often taken for granted by performing a comparative analysis between particle event lists and their solar origin.

\section{Description of different effects on the SEP events}

In this section we list various effects that influence the SEP events and provide, where possible, an estimation of their influence using primarily observational data.

\subsection{Physics-related data issues: Acceleration, injection and transport mechanisms}

Overall, particles can be energized by two physical processes in the solar atmosphere and interplanetary (IP) space, namely, by magnetic reconnection and shocks waves due to solar flares and CMEs (e.g., \cite{2010JGRA..11508101C,2013SSRv..175...53R}). The duration and efficiency of each acceleration process varies from event to event and depends on various reasons that are usually addressed by modeling (see recent review by \cite{2016LRSP...13....3D}).

\subsection{Location-related data issues: Magnetic connection between acceleration site and observer}

After the acceleration, escape and propagation through the IP medium, the SEP could be in principle detected by a satellite connected to the magnetic field line(s) guiding the particles. Namely, a magnetic connection must exist between the particles and the observer.

The particle instruments are situated in various locations: geostationary orbit, L1, or on a complicated route around Earth (as in the case for Wind spacecraft before 2004). This could be an additional reason for a given satellite not to observe a SEP event. 

Typically, SEP observations are done by a single spacecraft located at a single point in the heliosphere (usually in the ecliptic plane). At present, there is no possibility to reconstruct the SEP flux in the heliosphere based solely on observations. The observational limitation is the reason that only several hundred of particle events are recorded per solar cycle in contrast to the thousands of flare and CME events observed during the same period. Larger extents of the heliosphere can be monitored by suitably spaced spacecraft (e.g., the twin STEREO spacecraft \cite{2008SSRv..136....5K}, L5 (or/and L4) \cite{2016JASTP.146..171L} or/and out-of-ecliptic missions). These issues are intrinsic to the SEP phenomena or the detector location in space, respectively. Theoretical, numerical and remote-sensing efforts are needed in order to complement the lack of observational coverage. 

\subsection{Instrument-related data issues}

For the present study, we compared four data sets in the energy range $\sim$10 MeV, three proton lists in 25$-$30 MeV and four catalogs in the range 50$-$70 MeV. IMP-8 and the Wind/EPACT catalogs contain the most abundant data sets. A summary on the catalogs, instruments, time and energy coverage and number of events, used in this study is given in Table~\ref{T-catalogs}.

\begin{table}[t!]
\caption[]{Description of the proton catalogs used for the present comparative study. Abbreviations: re: re-calibrated data. The proton intensity range is given from min to max value. Status of the catalogs coverage as of April 2017.}
\small
\vspace{0.3cm}
\begin{tabular}{lllll}
\hline
Catalog name/   & Satellite/  & Temporal & Energy   & Intensity range\\
abbreviation    & Instrument/ & coverage & channel  & (number of events) \\
$[$reference$]$ & Data        &          & [MeV]    & in 1996$-$2006 \\
\hline
SEPEM$^a$ \cite{2015SpWea..13..406C}     & GOES/SEM$^{\rm re}$; & 1973$-$2013 & 7.23$-$10.45 & 0.53766$-$5505.5 \\
                                         & IMP-8           &             &              & (85)                  \\
GOES-NOAA$^b$                            & GOES/SEM        & 1976$-$2016 & $>$10        & 11$-$31700 (92)    \\
GOES-SSE \cite{2015SoPh..290..841D}      & SEPEM data      & 1997$-$2006 & $>$10        & 0.14$-$8200 (90)   \\
                                         &                 &             & $>$60        & 0.00051$-$950 (90) \\
GOES-SEP \cite{2016JSWSC...6A..42P}      & SEPEM data      & 1983$-$2013 & $>$10        & 2.38$-$25849.2 (135) \\
                                         &                 &             & $>$30        & 0.89$-$4253.31 (111) \\
                                         &                 &             & $>$60        & 0.55$-$861.13 (81) \\
IMP-8  \cite{2010JGRA..11508101C}        & IMP-8; SOHO     & 1997$-$2006 & $>$25        & 0.00001$-$45 (276) \\
Wind/EPACT$^c$                           & Wind/EPACT      & 1996$-$2016 & 19$-$28      & 0.0033$-$353.2 (280)  \\
\cite{2016simi.conf...27M,2017SunGeo..12.0M}  &            &             & 28$-$72      & 0.0003$-$32 (262) \\
SEPServer$^d$ \cite{2013JSWSC...3A..12V} & SOHO/ERNE       & 1997$-$2015 & 55$-$80      & 0.0006$-$0.5 (102)  \\
\hline
\multicolumn{5}{l}{$^a$ {\scriptsize \url{http://dev.sepem.oma.be/help/event_ref.html}}} \\
\multicolumn{5}{l}{$^b$ {\scriptsize \url{https://umbra.nascom.nasa.gov/SEP/}}} \\
\multicolumn{5}{l}{$^c$ {\scriptsize \url{http://newserver.stil.bas.bg/SEPcatalog/}}} \\
\multicolumn{5}{l}{$^d$ {\scriptsize \url{http://server.sepserver.eu/index.php?page=catalogue}}} \\
\end{tabular}
\label{T-catalogs}
\end{table}

\subsubsection{Data gaps}
Data gaps are expected to occur due to interruption of operation (satellite loss, instrument problem, or safe mode status), limited data transfer rate. One could identify when a SEP event is missed by comparing data from two independent instruments, and could recover (roughly estimate) the intensity of the missed proton event after performing a cross-correlation analysis between the fluxes of the two instruments.

\subsubsection{Sensitivity and dynamic range}
The possibility of detection of weak events by specific instrument is linked to the background intensity level during quiet times. For example, The Wind/EPACT instrument quiet background level is below 0.01 (cm$^2$ s sr MeV)$^{-1}$ for the low energy channel and below 0.001 for the high energy channel. Furthermore,  \cite{2013JSWSC...3A..12V} showed that the background level for the SOHO/ERNE instrument changes slightly with time epoch. 

The detectability limit for a given instrument considered here can be estimated by the lowest value of the peak SEP intensity ever observed, namely: 11 for GOES, 0.5377 for SEPEM, 0.00001 for IMP-8, 0.0033 for Wind/EPACT$^{i}$, 0.0003 for Wind/EPACT$^{h}$ and 0.0006 for SEPServer catalog (where GOES-based proton intensity is in proton flux unit (pfu = (cm$^2$ s sr)$^{-1}$) and the rest (IMP-8, Wind/EPACT and SEPServer) are in (cm$^2$ s sr MeV)$^{-1}$). 

A given detector may be with limited ability to record large proton intensities and thus will suffer from saturation during large events. For example, for about 11\% (12/112) of the reported proton events by the SEPServer catalog only a lower limit for their peak flux can be given due to saturation of the energy channels. Since these events should be dropped from quantitative analysis the catalog is biased to low-intensity SEP events.

\subsubsection{Contamination}
Occasionally, flux from other species and/or energy channels (e.g., pile-up effect) can enter the detector and be recorded erroneously with limited possibilities for correction. Such cases have to be excluded from further scientific analysis since the particle flux measurements are unreliable.

\subsubsection{Energy range}
Different satellites observe in different energy channels. The chosen catalogs are roughly divided into low ($\sim$10 MeV), intermediate (19$-$28 and $>$30 MeV) and high (with detectors over 28$-$72, $>$60 and 55$-$80 MeV) energy range. The energy channels largely overlap each other, however the coverage is not complete. In order to provide a quantitative evaluation of the possible differences, in this study we present the cross-correlation between the different catalogs (see Section~3.1).

\subsection{Observer-based data issues}

\subsubsection{Selection effects, subjectivities and assumptions}
Visual identification of SEP events is a common start point for the selection of particle enhancement for many catalogs. For manual catalogs, observers scan the intensity$-$time plots and visually identify a proton enhancement (as done for the GOES-SSE and Wind/EPACT lists). The SEPServer team reports a $\sim$68 MeV event when the 1-min average proton intensity is enhanced $\sim$3 times over the quiet-time background level. Constant proton intensity threshold levels are imposed in the case of automatic procedures. Namely, the GOES-NOAA list requires the 10 pfu flux level to be surpassed in order to report a proton event of $>$10 MeV energy range. Either approach has limitations (missed events, erroneous identifications) and advantages (e.g., the automatic regime can trigger alerts in near real time). For the case of the GOES-NOAA proton list, the selected threshold for reporting a new SEP event led to numerous omissions. We evaluated more than 100 weak events (below the 10 pfu threshold level) that are not reported by GOES-NOAA catalog, some included in the GOES-SEP list. Thus, the GOES-NOAA list is biased to strong events starting from a (pre-event) quiet-time intensity level which for GOES-NOAA data is usually $\lesssim$1 pfu. In addition, if a new SEP increase occurs when the flux level is still above the 10 pfu (i.e., a new SEP injection in the aftermath of a strong, ongoing particle event), the new proton event will not be reported. We estimated that about 10 events are omitted due to this reason. The re-analysis of the GOES-NOAA proton events (independently on the available GOES-SEE and GOES-SEP catalogs) is currently under completion and will be reported elsewhere.

An additional subjectivity is imposed when a number of SEP events is not considered for the aim of the specific scientific goal. Often, only large SEP events are taken into account, e.g., \cite{2004JGRA..10912105G} selects about 30 major events for their study. In this work we also investigate the effect of selecting different subsets based on proton intensity.

For the analysis in the present study, we compared SEP events from seven different proton catalogs, see Table~\ref{T-catalogs}. In summary, we identified 361 proton events observed at Earth during SC23.

\subsubsection{Onset time definitions}
There is no unique way to evaluate the time of the SEP onset. Different definitions are still in use, e.g., SEP onset is defined when the particle flux: rises above pre-event background level of the amount of: 2 \cite{2003ICRC....6.3305T}, 3 \cite{2017SunGeo..12.0M}, or 4 \cite{1999ApJ...519..864K} standard deviations (sigmas); surpasses a fixed intensity level (e.g., 10 pfu threshold adopted for the GOES proton catalog); is calculated by the Poisson-CUSUM method \cite{2005A&A...442..673H}; is evaluated using the intersection point between the background and the fitting line to the particle profile \cite{2014SoPh..289.2601M}. With the exception of the latter case, no error margins are given. If some smoothing on the data is performed, the results will change and the uncertainty on the onset time determination will increase.

\subsubsection{Solar origin identifications}
The origin of SEP events is usually evaluated by associating a flare and CME (the so-called SEP origin) prior the SEP release time. Usually, the pair with largest flare class/fastest CME is chosen but the procedure includes subjectivity. The proton release time is not directly observed but is being deduced by the observed SEP onset time. If the approximation of a scatter-free transport is adopted, one can estimate the travel time for a given particle speed and obtain the latest possible injection time (of SEPs leaving the solar corona). Using such timing arguments, the SEP solar source is proposed. There are several catalogs for flare and CME characteristics that can be used. The parameters of solar flares are collected from the GOES soft X-ray (SXR) instrument reports available on-line by several sources: flare listings by NOAA\footnote{\url{ftp://ftp.ngdc.noaa.gov/STP/space-weather/solar-data/solar-features/solar-flares/x-rays/goes/xrs/}}; NASA\footnote{\url{http://hesperia.gsfc.nasa.gov/goes/goes_event_listings/}}; preliminary and comprehensive reports of the solar geophysical data\footnote{\url{http://www.ngdc.noaa.gov/stp/space-weather/online-publications/stp_sgd/}}; Solar Monitor\footnote{http://www.solarmonitor.org}. The CME properties are provided by: SOHO/LASCO CDAW\footnote{\url{http://cdaw.gsfc.nasa.gov/CME_list/}} \cite{2009EM&P..104..295G}); CACTus\footnote{\url{http://sidc.be/cactus/}}; SEEDS\footnote{\url{http://spaceweather.gmu.edu/seeds/}}; CORIMP\footnote{\url{http://alshamess.ifa.hawaii.edu/CORIMP/}} catalogs, respectively. The discrepancies between the different flare databases are currently being explored and will be reported elsewhere. The offsets in the CME reports provided by CDAW and CACTus are discussed in \cite{2009ApJ...691.1222R}. A recent study by \cite{2015SoPh..290.1741R} compares the correlations between the proton intensity and CME speed when using various CME catalogs and shows that the results are consistent.

\subsubsection{Local particle acceleration}
We completed a visual inspection of the GOES proton enhancements and estimated about 30\% of energetic storm particle (ESP) contributions that were reported as SEP events by the GOES-NOAA proton database. Namely, the GOES-NOAA catalog reports erroneously elevated peak values for these cases. However, \cite{2015SoPh..290..841D} (based on the GOES-SSE list) showed that in terms of linear correlation analysis with flare/CME parameters, the SEP and ESP intensities give consistent results within the uncertainties. In the present study we used the values from the GOES-NOAA, GOES-SSE and GOES-SEP catalogs as reported.

\subsubsection{Projection effects: flares}
Occasionally, no flare can be identified as the SEP origin. In the majority of these cases, the flare is behind the limb, as inspected by EUV images and by occulted radio emission signatures. In addition, the SXR emission of limb flares is partially occulted and only a lower limit for the flare intensity is given in the flare listings. This introduces a bias when doing correlation between the observed particle and flare intensities. Some estimation for the occulted and back-sided flare emission is reported by \cite{2013SoPh..288..241N} and \cite{2015SoPh..290.1947C}. However the number of events provided by these databases are not sufficient for the purpose of large statistical studies. In addition, there are no estimations before the STEREO era.

\subsubsection{Projection effects: CMEs}

The reported CME speed is the projected component on the plane of the sky. In overall, the reported (projected) CME speed is an underestimation. Limb events, with radially outward propagation, will be the least affected.

There are different schemes proposed to reconstruct the 3D speed of CMEs, e.g., \cite{2008JGRA..113.1104H} and \cite{2013ApJ...777..167D}. Different sets of assumptions are employed. Recent attempts to build a database of the de-projected CME speeds was announced by the HELCATS project\footnote{https://www.helcats-fp7.eu/}. The time coverage after 2007 is the reason not to consider this database for the purpose of statistical study here.

\subsubsection{Statistical approaches: limitations and uncertainties}
The solar origin identification (flare or CME) is further evaluated based on a strength of the linear correlation between the peak proton intensity with the flare class or with the CME projected speed\footnote{Note that the widely adopted procedure to select an event-integrated or fixed in time SEP/flare/CME parameter in order to represent the complex SEP/flare/CME process, respectively, is a gross approximation.}. While comparing the obtained two correlation coefficients, often no uncertainties are evaluated and the data samples are rather small (often well below 100s of events). Small differences (below 20\%) in the correlations are used to argue in favor of one or the other accelerator. For example, \cite{2003GeoRL..30lSEP3G} and \cite{1982ApJ...261..710K} reported correlations of 0.4 to 0.5 with the flare, whereas stronger correlations with the CMEs were given by \cite{2003GeoRL..30lSEP3G} and \cite{2001JGR...10620947K}, 0.6 and 0.7, respectively. Others \cite{2010JGRA..11508101C,2013SoPh..282..579M} evaluated the same correlation ($\sim$0.6) with both accelerators. Recently, \cite{2013SoPh..282..579M} implemented the bootstrapping method \cite{2003psa..book.....W} and showed that the statistical uncertainty can reach of up to 20\%. The uncertainty tends to decrease when the data scatter is small but also when the data sample is large. In this work, the same method is used to evaluate the uncertainty of the correlations.

\section{Results}

\subsection{Comparison between different catalogs}

\begin{figure}[!t]
\centerline{\includegraphics[width=\textwidth,clip=]{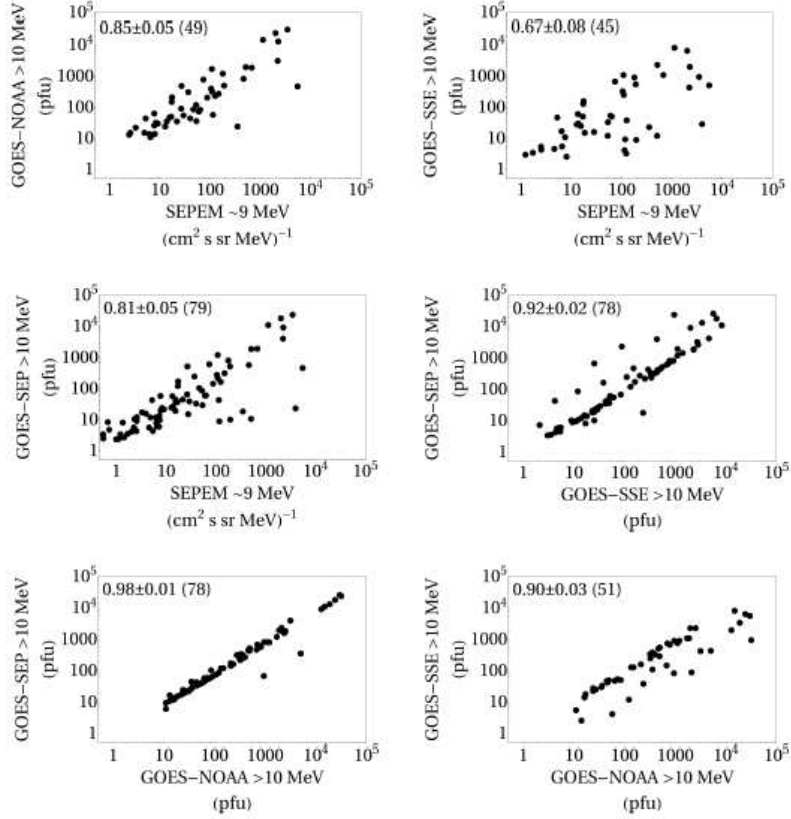}}       
\caption[]{Correlation plots (log$-$log) between the peak proton intensity at $\sim$10 MeV in SC23 (1996$-$2006) from SEPEM, GOES-NOAA, GOES-SSE and GOES-SEP catalogs. The correlation coefficient, its uncertainty and exact number of events (in brackets) are given on each scatter plot.}
    \label{F-Scatterplot10}
\end{figure}

\begin{figure}[!t]
\centerline{\includegraphics[width=0.7\textwidth,clip=]{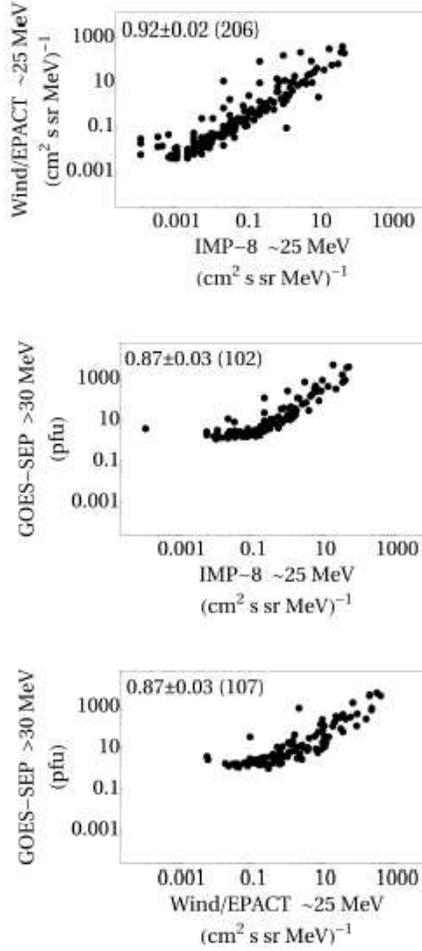}}       
\caption[]{Correlation plots (log$-$log) between the peak proton intensity at $\sim$25 MeV in SC23 (1996$-$2006) from IMP-8, Wind/EPACT and GOES-SEP catalogs. The correlation coefficient, its uncertainty and exact number of events (in brackets) are given on each scatter plot.}
    \label{F-Scatterplot25}
\end{figure}

\begin{figure}[!t]
\centerline{\includegraphics[width=0.95\textwidth,clip=]{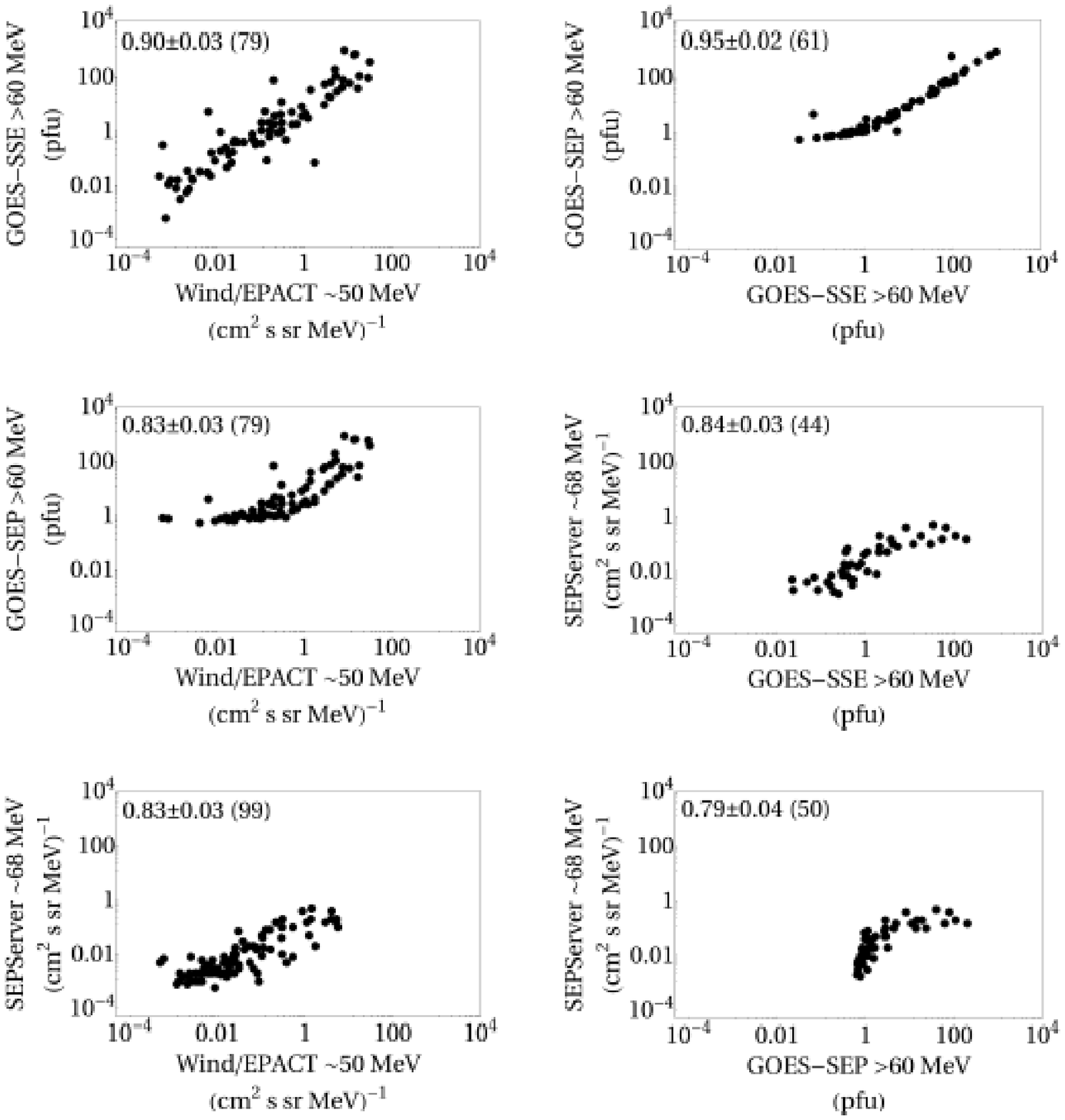}}   
\caption[]{Correlation plots (log$-$log) between the peak proton intensity at $\sim$60 MeV in SC23 (1996$-$2006) from Wind/EPACT, GOES-SSE, GOES-SEP and SEPServer catalogs. The correlation coefficient, its uncertainty and exact number of events (in brackets) are given on each scatter plot.}
    \label{F-Scatterplot60}
\end{figure}

We identified the same SEP event in the different particle lists using the following criteria. The reported proton onset times are within one day, the reported peak intensities are both larger/smaller than the median value of the sample, and the associated flares/CMEs (if provided) are the same. At least two of these requirements should apply. Following this procedure we identified 361 individual proton events. About 32\% (116/361) are observed by a single spacecraft, whereas the rest are observed by two and more satellites (and only $\sim$4\% are reported by all seven catalogs). We found that the time offset (between the reported onset times) range from 10s minutes to several hours. Information on the catalogs, instruments, energy and intensity ranges (minimum to maximum value) and the final number of proton events in the considered time period are summarized in Table~\ref{T-catalogs}. We organize the comparative study based on the proton energy, namely low (`l') at average energy of $\sim$10 MeV, intermediate (`i') at $\sim$25 MeV and high (`h') at $\sim$60 MeV, as introduced above.

A cross-correlation was performed on the SEP events as reported by the catalog pairs during SC23, see Figs.~\ref{F-Scatterplot10}$-$\ref{F-Scatterplot60}. The Pearson correlation coefficients (log$-$log) are calculated for each catalog pair and given in each figure. We obtained the highest correlation for GOES-SEP and GOES-NOAA (0.98$^{\pm 0.01}$) and the lowest for GOES-SSE and SEPEM (0.67$^{\pm 0.08}$). For the intermediate energies (Fig.~\ref{F-Scatterplot25}) the correlation is higher than 0.87 and at high energies (Fig.~\ref{F-Scatterplot60}) the correlation is higher than 0.79. The number of SEP events common for each pair of catalogs is given in brackets in each plot. For completeness we calculated the cross-correlation for the other combinations, however the scatter there is larger, and the correlations decrease to 0.45.

Overall, the results show a consistency between proton intensity trends observed by a given instrument compared to other instruments with similar energy coverage, with correlation coefficients ranging from 0.67 to 0.98. The larger scatter on some of the plots could be due to the difference in energy channels, an erroneous identification of the same SEP event in the specific catalog pair (e.g., due to insufficient precision of the reported onset timing), due to the different intensity units used (pfu and (cm$^2$ s sr MeV)$^{-1}$), or other instrument issues. 

\subsection{Temporal trends}

\begin{figure}[!t]
\centerline{\includegraphics[width=0.75\textwidth,clip=]{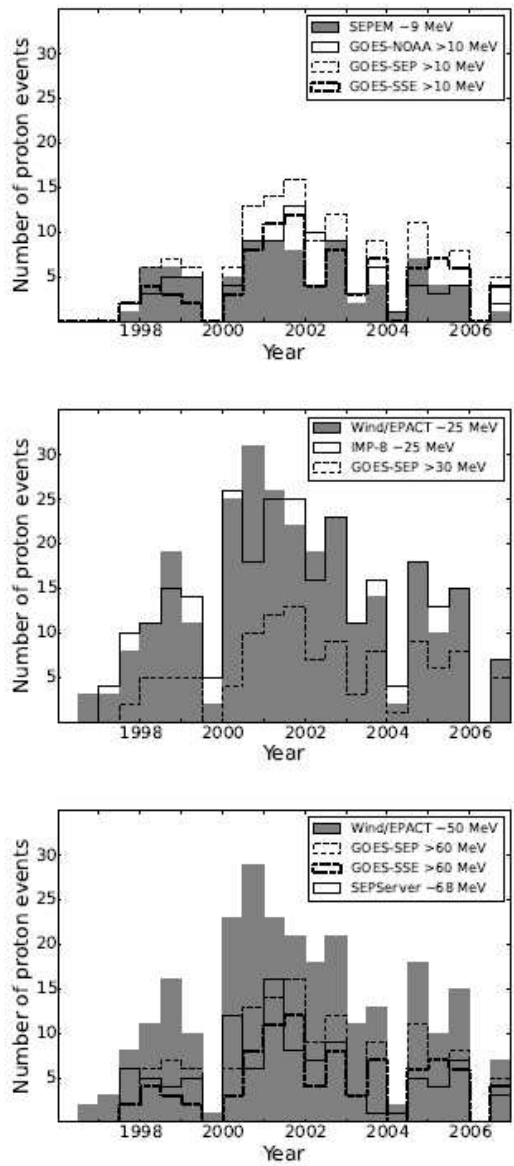}}
\caption[]{Temporal distribution of the SEP events from each dataset during the period 1996$-$2006 using 6-month binning.}
    \label{F-timetrends}
\end{figure}

In Figure~\ref{F-timetrends} we show the distribution of SEP events over 1996$-$2006, given separately for each energy range. The distributions are consistent. All SEP samples peak between 2000$-$2002. A slight increase in the number of events is noticeable at the declining phase of SC23 whereas all catalogs show a localized event minimum at around 2004.

\subsection{Different onset time definitions}

Onset times are reported for all SEP catalogs. For the case of IMP-8 data set the reported onset times are given rounded to the nearest hour which is in general an insufficient accuracy (and thus the uncetainty is at least one hour). The accuracy for the Wind/EPACT catalog is about 7.5 min due to the smoothing used. The integrated GOES proton fluxes are 5-minute averages. For the SEPServer catalog we adopt an accuracy of one minute due to the reported precision of the onset time. 

We obtained the following temporal differences between the reported SEP onset times listed below in terms of mean/median values (and given in hours):
\begin{itemize}
 \item SEPEM$-$GOES-NOAA: $-$7.5/$-$2.8
 \item GOES-SSE$^l-$GOES-NOAA: $-$3/$-$1.3
 \item GOES-SEP$^l-$GOES-NOAA: $-$5.4/$-$2.3 
 \item SEPEM$-$GOES-SSE$^l$: $-$2/0
 \item SEPEM$-$GOES-SEP$^l$: +1.5/+1.1
 \item GOES-SSE$^l-$GOES-SEP$^l$: +1.3/+0.4
 \item Wind/EPACT$^i-$IMP-8: +2.3/+1.7
 \item Wind/EPACT$^h-$SEPServer: +0.8/+0.5
\end{itemize}

Negative values denote that the SEP onset times from the first data set occur before the values from the second catalog and vice versa. We obtain that SEPEM, GOES-SSE$^l$ and GOES-SEP$^l$ report similar onset times (within about one hour), whereas they all are systematically earlier than the onset times evaluated for GOES-NOAA catalog (using fixed intensity threshold). The Wind/EPACT$^i$ and Wind/EPACT$^h$ onsets are slightly later in time than the reported onsets by IMP-8 list and SEPServer catalog (using the CUSSUM method), respectively, for the same SEP event.

\subsection{Correlation coefficients with flares and CMEs}

\begin{figure}[!t]
\centerline{\includegraphics[width=\textwidth,clip=]{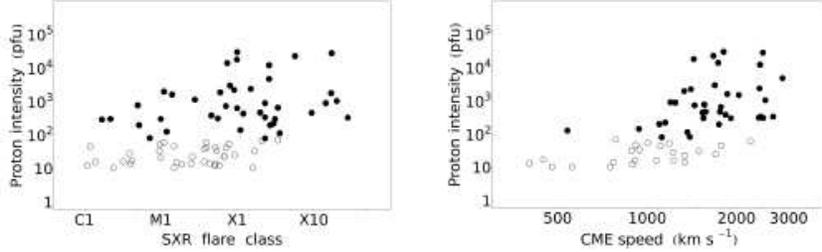}}  
\caption[]{Correlation plots (log$-$log) between the peak proton intensity from GOES-NOAA catalog and the flare class (left plot) and CME speed (right). Open/filled circles are for small/large protons with respect to the median intensity.}
    \label{F-Examples}
\end{figure}

We calculate the Pearson correlation between the log$_{10}$ values of the peak proton intensity and log$_{10}$ of the flare class and/or CME projected speed. Examples of these so-called log$-$log scatter plots are shown in Figure~\ref{F-Examples} for the GOES-NOAA catalog. With different symbols we show the protons with peak intensity above (filled) and below (open circles) the median intensity value for the sample. The entire proton sample is the sum of both cases. Adopting a specific proton intensity threshold, e.g., considering only large (also termed major) proton events, will change the correlation coefficients.

In Table~\ref{T-cc} we summarize the correlation coefficients in SC23 between all reported proton events from the different catalogs with the identified solar origin. Additionally, we do the same calculations using the large proton events (namely with intensity larger than the median value for the sample). We calculated the Pearson correlations (log$-$log) whereas the uncertainties are estimated using the bootstrapping method. 

For the statistical analysis here, we identified the SEP solar origin (flare and CME) as discussed in Section~2.4.3, and in \cite{2017SunGeo..12.0M}. From the entire SEP event list (361 events), we could identify 256/361 (71\%) SEP-associated flares and 277/361 (77\%) SEP-associated CMEs. The reasons for the lower number of the solar origin events are: data gaps (often for the CMEs), multiple eruptions (complex cases), back-sided events (usually for the flares), and uncertain cases.

\begin{table}[t!]
\caption[]{Table with log$-$log correlation coefficients and their uncertainties calculated between the SEP peak intensity for the all and large proton events with SF class or CME speed, respectively, in the period 1996$-$2006. No restriction of flare class or CME speed is imposed. The number of events used to calculate the coefficients is given in brackets. Abbreviations: {\it l}: low energy; {\it i}: intermediate energy; {\it h}: high energy.}
\label{T-cc}
\small
\vspace{0.3cm}
\begin{tabular}{lllll}
\hline
Proton         & \multicolumn{4}{c}{Correlation coefficients}\\
event          & \multicolumn{2}{c}{all proton events with:} & \multicolumn{2}{c}{large proton events with:}  \\
sample         & SF class & CME speed & SF class & CME speed \\
\hline
SEPEM          & $0.37^{\pm 0.10}$ (69) & $0.31^{\pm 0.10}$ (68) & $0.21^{\pm 0.15}$ (39) & $0.30^{\pm 0.20}$ (33) \\
GOES-NOAA      & $0.47^{\pm 0.07}$ (78) & $0.52^{\pm 0.06}$ (76) & $0.24^{\pm 0.12}$ (41) & $0.38^{\pm 0.12}$ (38) \\
GOES-SSE$^l$   & $0.53^{\pm 0.08}$ (88) & $0.54^{\pm 0.07}$ (79) & $0.39^{\pm 0.12}$ (45) & $0.38^{\pm 0.14}$ (40) \\
GOES-SEP$^l$   & $0.45^{\pm 0.07}$ (116)& $0.49^{\pm 0.05}$ (116)& $0.38^{\pm 0.09}$ (62) & $0.40^{\pm 0.09}$ (57) \\
\hline
IMP-8          & $0.47^{\pm 0.06}$ (222)& $0.42^{\pm 0.06}$ (231)& $0.43^{\pm 0.08}$ (122)& $0.48^{\pm 0.07}$ (123) \\
Wind/EPACT$^i$ & $0.43^{\pm 0.06}$ (202)& $0.51^{\pm 0.05}$ (220)& $0.39^{\pm 0.08}$ (106)& $0.44^{\pm 0.07}$ (109) \\
GOES-SEP$^i$   & $0.46^{\pm 0.08}$ (98) & $0.42^{\pm 0.06}$ (97) & $0.44^{\pm 0.09}$ (53) & $0.25^{\pm 0.11}$ (47) \\
\hline
Wind/EPACT$^h$ & $0.46^{\pm 0.06}$ (193)& $0.50^{\pm 0.05}$ (209)& $0.42^{\pm 0.08}$ (100)& $0.41^{\pm 0.08}$ (104) \\
GOES-SSE$^h$   & $0.59^{\pm 0.07}$ (88) & $0.42^{\pm 0.07}$ (79) & $0.40^{\pm 0.12}$ (45) & $0.17^{\pm 0.16}$ (38) \\
GOES-SEP$^h$   & $0.46^{\pm 0.09}$ (74) & $0.35^{\pm 0.09}$ (71) & $0.33^{\pm 0.12}$ (40) & $0.17^{\pm 0.18}$ (34) \\
SEPServer      & $0.39^{\pm 0.11}$ (86) & $0.32^{\pm 0.09}$ (94) & $0.22^{\pm 0.14}$ (48) & $0.09^{\pm 0.14}$ (47) \\
\hline
\end{tabular}
\end{table}

The value of the correlation coefficients between the proton intensity and the flare class in the low energy channels ranges between 0.37$^{\pm 0.10}$ and 0.59$^{\pm 0.07}$ that largely overlays the range of the correlations with the CME projected speed that has a value as low as 0.31$^{\pm 0.10}$ (see Table~\ref{T-cc}). The ranges for the other energy channels are similar. The error bars range from 5 to 11\%. In general, the differences in the correlation coefficients (between the entire proton sample with with flares or with CMEs) are within the uncertainties. These estimates are for all cases in SC23, where flare and CME could be identified, irrelevant on the location of their parent active region. No clear decreasing/increasing trend of the correlation coefficients with proton energy is evident.

In addition, we investigate the change in correlation coefficients when considering large intensity SEP events. The latter is a common start point for many earlier studies. Namely, we aim to investigate the source of the reported discrepancies in the correlations when using the entire and a sub-set of the catalogs. We selected a sub-set from each catalog containing events with peak proton intensity larger than the median value for the given dataset. In Table~\ref{T-cc} we present the results calculated following similar procedure as for the entire proton sample.

While comparing the corresponding columns of the table we note that overall the coefficients between large SEP events and flares/CMEs are lower in values that the corresponding coefficients considering the entire proton sample with flares/CMEs. At low energy, the correlations between large proton events with flares are lower than the respective correlations with CMEs, as reported previously. In general, the opposite is obtained for the intermediate and high energies. The decreasing/increasing tendencies are not statistically significant, since the correlations based on the sample of large proton events have large error bars, mostly due to fewer number of events there. Based on the provided statistical uncertainty the significance of the results could be evaluated.

\subsection{Projection effects}

Both, the flare SXR class and the CME speed are subject to projection effects. Flares close to the limb can be partially occulted. A number of SEP-associated flares probably occurred on the back-side and these cases are dropped from the correlation analysis since no estimate of the SXR flare intensity is possible. The value for the CME speed used here is the reported projected component.

\begin{table}[t!]
\caption[]{Table with log$-$log correlation coefficients and their uncertainties calculated between the SEP peak intensity for the all and large proton events with SF class or CME speed, respectively, in the period 1996$-$2006. We select on-disk flares, with origin $\leq$45 degrees, and limb CMEs, with origin $>$45 degrees helio-longitude, respectively. The number of events used to calculate the coefficients is given in brackets. Abbreviations: {\it l}: low energy; {\it i}: intermediate energy; {\it h}: high energy.}
\label{T-cc-proj}
\small
\vspace{0.3cm}
\begin{tabular}{lllll}
\hline
Proton         & \multicolumn{4}{c}{Correlation coefficients}\\
event          & \multicolumn{2}{c}{all proton events with:} & \multicolumn{2}{c}{large proton events with:}  \\
sample         & SF class & CME speed & SF class & CME speed \\
\hline
SEPEM          & $0.27^{\pm 0.18}$ (32) & $0.27^{\pm 0.15}$ (27) & $-0.02^{\pm 0.19}$ (22)& $0.47^{\pm 0.21}$ (13) \\
GOES-NOAA      & $0.36^{\pm 0.15}$ (31) & $0.48^{\pm 0.08}$ (39) & $0.29^{\pm 0.20}$ (21) & $0.22^{\pm 0.16}$ (20) \\
GOES-SSE$^l$   & $0.54^{\pm 0.11}$ (46) & $0.48^{\pm 0.13}$ (35) & $0.50^{\pm 0.15}$ (22) & $0.21^{\pm 0.22}$ (19) \\
GOES-SEP$^l$   & $0.53^{\pm 0.10}$ (56) & $0.38^{\pm 0.11}$ (50) & $0.31^{\pm 0.16}$ (29) & $0.28^{\pm 0.17}$ (26) \\
\hline
IMP-8          & $0.61^{\pm 0.08}$ (101)& $0.36^{\pm 0.09}$ (97) & $0.64^{\pm 0.08}$ (58) & $0.34^{\pm 0.13}$ (50) \\
Wind/EPACT$^i$ & $0.60^{\pm 0.07}$ (86) & $0.49^{\pm 0.08}$ (93) & $0.56^{\pm 0.09}$ (46) & $0.35^{\pm 0.13}$ (47) \\
GOES-SEP$^i$   & $0.54^{\pm 0.09}$ (45) & $0.28^{\pm 0.11}$ (44) & $0.57^{\pm 0.13}$ (24) & $0.18^{\pm 0.23}$ (22) \\
\hline
Wind/EPACT$^h$ & $0.59^{\pm 0.07}$ (81) & $0.48^{\pm 0.08}$ (91) & $0.60^{\pm 0.09}$ (42) & $0.28^{\pm 0.14}$ (45) \\
GOES-SSE$^h$   & $0.62^{\pm 0.09}$ (46) & $0.31^{\pm 0.13}$ (35) & $0.63^{\pm 0.14}$ (20) & $0.04^{\pm 0.26}$ (19) \\
GOES-SEP$^h$   & $0.55^{\pm 0.10}$ (36) & $0.28^{\pm 0.15}$ (30) & $0.56^{\pm 0.20}$ (16) & $-0.05^{\pm 0.30}$ (18) \\
SEPServer      & $0.55^{\pm 0.10}$ (37) & $0.28^{\pm 0.14}$ (41) & $0.40^{\pm 0.19}$ (18) & $0.12^{\pm 0.21}$ (24) \\
\hline
\end{tabular}
\end{table}

In order to minimize the projection effects on the correlations, in the present study we dropped the cases expected to be severely affected by projection effects, namely flares located close to the solar limb and CMEs erupting close to the solar disk center. There is some subjectivity while selecting the border between on-disk and limb events. Here, we adopt the value of 45 degrees. We present the coefficients for on-disk flares (located from E45 to W45) and limb CMEs (at helio-longitude $>$E45 and $>$W45 based on the location of the AR of the CME-associated flare). All correlations coefficients are listed in the respective columns in Table~\ref{T-cc-proj}.

With the exception of the 10 MeV SEPEM and GOES-NOAA datasets, there is an increasing tendency, compared to Table~\ref{T-cc} for the correlations between all protons with on-disk flares, as expected, since the SXR emission for all flares is expected to be reduced due to partial occultation. For the case of CMEs, the removal of events erupting close to the solar disk center does not improve the correlations, but actually reduces slightly all values. For the case of large protons, in general, both increasing and decreasing tendencies are obtained compared to the large protons with no restriction on flares/CMEs (from Table~\ref{T-cc}). Due to the large uncertainties, however, the differences between all and the reduced event sample are mostly not statistically significant. 

\section{Summary and discussion}

In the present study we summarized various effects that influence the observed in situ SEP intensity time profiles in terms of physical, instrumental and observer aspects. The list of possible effects could be used as a set of guidelines for particle detection and for the interpretations of the particle data. In the present work we quantify only selected aspects on SEP studies using observational data and different techniques, as given below.
 
There is a different number of proton events reported by the different event catalogs even when the instruments cover similar energy and ranges and time periods. Possible explanations for this reason are data gaps (e.g., during the period of SOHO loss), instrument sensitivity (e.g., 165 more events are reported by IMP-8 $\sim$25 MeV, compared to GOES-SEP $>$30 MeV), and the adopted definition for SEP event identification (e.g., 44 more events are reported by GOES-SEP compared to GOES-NOAA, for the same instrument, energy range and time period).
 
We found that fixed threshold in particle intensity (as for the GOES-NOAA catalog) give systematically later SEP onset times (by hours) which proves to be as the least accurate method used. Fixes SEP intensity levels (e.g., large SEP events) also introduces a bias when performing correlation coefficients, namely towards reduced correlation with flare class and enhanced correlation with CME speed. This could be one of the explanation of the reported differences in correlation coefficients by earlier works (see comparison done in \cite{2013CEAB...37..541M}). Other reasons include different identifications for the solar origin (flares/CMEs) of the same proton event, energy dependence effect (see trends found by \cite{2015SoPh..290..841D}), different flare/CME databases used.

Large proton events were shown in earlier studies, based on GOES-NOAA catalog, to have higher correlations with CMEs compared to flares \cite{2003GeoRL..30lSEP3G,2014EP&S...66..104G}. No error bars on the coefficients, however, were reported. In our analysis, when we select a sub-sample from larger than median value proton events, we obtain a similar tendency for SEPEM and GOES-NOAA at $\sim$10 MeV and for IMP-8 and Wind/EPACT $\sim$25 MeV. At higher energy, e.g. for GOES-SSE $>$60 MeV, GOES-SEP $>$60 MeV and SEPServer catalogs, we obtain the opposite result. Neither of these tendencies are statistically significant, however, which impressively demonstrates the need for validation of the performed statistics. Proton energy and abundance of the event sample under consideration are also to be considered before doing a solar origin interpretation.

Projection effects are expected to reduce the reported flare class and CME speed and thus the correlations of protons with the de-projected values for flare class and CME speed could change. In the present study, we aimed to reduce the projection effects by applying the exclusion principle in contrast to adopting certain method for correction. On-disk and limb events are approximately separated here at the 45 degree heliolongitude. We confirm an increasing tendency when calculating the Pearson coefficients between the SEP peak flux and on disk flare class. The GOES-NOAA and SEPEM $\sim$10 MeV databases show a decline that could be due to the methodology for their identification. For the case of limb CMEs, however, we obtain a slight reduction in the correlation coefficients. One possible explanation is that based on the here-chosen criteria for limb CMEs, the sub-set still contains a numerous CME events that suffer by large projection effects. This possibility, added to the overall reduced number of the sample lead to overall reduction of the obtained correlations.

The solar origin proposed by the different teams is not identical but varies to certain degree. This could be another reason for the differences in the reported correlation coefficients. For example, for the IMP-8 data in this work we obtain below 0.5 with flares and above 0.4 with CMEs, whereas, over the same period, \cite{2010JGRA..11508101C} reports correlations of $\sim$0.6 for both. When considering other catalogs (e.g., GOES-SEP), the different time period under consideration (the different number of events in the samples) could be an additional reason for the difference in the reported correlations.

In summary, our results confirm there are number of reasons of different origin that need to be considered when performing SEP observations, identification, statistical analysis and their interpretation. Multiple effects influence the observed results that cannot be corrected for, e.g., location of satellite, data gaps, particle propagation, subjectivity while selecting the SEP origin. On the other side, any particle study should provide the correct identification of SEP events recorded by a given instrument (in contrast to ESP signatures reported as SEPs), onset times and peak intensities with their uncertainty ranges, as well as validated statistical analysis. Finally, improved theoretical models are expected to minimize the limitations of present day observations.

\section*{Acknowledgments}
We acknowledge the freely available data and catalogs for SEP, flare and CME events. The CME catalog is generated and maintained at the CDAW Data Center by NASA and The Catholic University of America in cooperation with the Naval Research Laboratory. SOHO is a project of international cooperation between ESA and NASA. MVCD thanks FAPESP project No. 2016/05254-9.

\section*{References}

\bibliography{mybibfile}

\begin{thebibliography}{10}
\expandafter\ifx\csname url\endcsname\relax
  \def\url#1{\texttt{#1}}\fi
\expandafter\ifx\csname urlprefix\endcsname\relax\def\urlprefix{URL }\fi
\expandafter\ifx\csname href\endcsname\relax
  \def\href#1#2{#2} \def\path#1{#1}\fi

\bibitem{2016LRSP...13....3D}
M.~{Desai}, J.~{Giacalone}, {Large gradual solar energetic particle events},
  Living Reviews in \solphys 13 (2016) 3.
\newblock \href {http://dx.doi.org/10.1007/s41116-016-0002-5}
  {\path{doi:10.1007/s41116-016-0002-5}}.

\bibitem{2008SSRv..136....5K}
M.~L. {Kaiser}, T.~A. {Kucera}, J.~M. {Davila}, O.~C. {St.~Cyr},
  M.~{Guhathakurta}, E.~{Christian}, {The STEREO Mission: An Introduction},
  \ssr 136 (2008) 5--16.
\newblock \href {http://dx.doi.org/10.1007/s11214-007-9277-0}
  {\path{doi:10.1007/s11214-007-9277-0}}.

\bibitem{1992A&AS...92..365S}
J.~A. {Simpson}, J.~D. {Anglin}, A.~{Balogh}, M.~{Bercovitch}, J.~M. {Bouman},
  E.~E. {Budzinski}, J.~R. {Burrows}, R.~{Carvell}, J.~J. {Connell},
  R.~{Ducros}, P.~{Ferrando}, J.~{Firth}, M.~{Garcia-Munoz}, J.~{Henrion},
  R.~J. {Hynds}, B.~{Iwers}, R.~{Jacquet}, H.~{Kunow}, G.~{Lentz}, R.~G.
  {Marsden}, R.~B. {Mckibben}, R.~{Meuller-Mellin}, D.~E. {Page}, M.~{Perkins},
  A.~{Raviart}, T.~R. {Sanderson}, H.~{Sierks}, L.~{Treguer}, A.~J.
  {Tuzzolino}, K.~P. {Wenzel}, G.~{Wibberenz}, {The ULYSSES Cosmic Ray and
  Solar Particle Investigation}, \aaps 92 (1992) 365--399.

\bibitem{1995SoPh..162....1D}
V.~{Domingo}, B.~{Fleck}, A.~I. {Poland}, {The SOHO Mission: an Overview},
  \solphys 162 (1995) 1--37.
\newblock \href {http://dx.doi.org/10.1007/BF00733425}
  {\path{doi:10.1007/BF00733425}}.

\bibitem{1995SoPh..162..505T}
J.~{Torsti}, E.~{Valtonen}, M.~{Lumme}, P.~{Peltonen}, T.~{Eronen},
  M.~{Louhola}, E.~{Riihonen}, G.~{Schultz}, M.~{Teittinen}, K.~{Ahola},
  C.~{Holmlund}, V.~{Kelh{\"a}}, K.~{Lepp{\"a}l{\"a}}, P.~{Ruuska},
  E.~{Str{\"o}mmer}, {Energetic Particle Experiment ERNE}, \solphys 162 (1995)
  505--531.
\newblock \href {http://dx.doi.org/10.1007/BF00733438}
  {\path{doi:10.1007/BF00733438}}.

\bibitem{1995SSRv...71..155V}
T.~T. {von Rosenvinge}, L.~M. {Barbier}, J.~{Karsch}, R.~{Liberman}, M.~P.
  {Madden}, T.~{Nolan}, D.~V. {Reames}, L.~{Ryan}, S.~{Singh}, H.~{Trexel},
  G.~{Winkert}, G.~M. {Mason}, D.~C. {Hamilton}, P.~{Walpole}, {The Energetic
  Particles: Acceleration, Composition, and Transport (EPACT) investigation on
  the WIND spacecraft}, \ssr 71 (1995) 155--206.
\newblock \href {http://dx.doi.org/10.1007/BF00751329}
  {\path{doi:10.1007/BF00751329}}.

\bibitem{2014SW...12..92R}
J.~V. {Rodriguez}, J.~C. {Krosschell}, J.~C. {Green}, {Intercalibration of GOES
  8–15 solar proton detectors}, Space Weather 12.

\bibitem{2015SpWea..13..406C}
N.~{Crosby}, D.~{Heynderickx}, P.~{Jiggens}, A.~{Aran}, B.~{Sanahuja},
  P.~{Truscott}, F.~{Lei}, C.~{Jacobs}, S.~{Poedts}, S.~{Gabriel},
  I.~{Sandberg}, A.~{Glover}, A.~{Hilgers}, {SEPEM: A tool for statistical
  modeling the solar energetic particle environment}, Space Weather 13 (2015)
  406--426.
\newblock \href {http://dx.doi.org/10.1002/2013SW001008}
  {\path{doi:10.1002/2013SW001008}}.

\bibitem{2010JGRA..11508101C}
H.~V. {Cane}, I.~G. {Richardson}, T.~T. {von Rosenvinge}, {A study of solar
  energetic particle events of 1997-2006: Their composition and associations},
  Journal of Geophysical Research (Space Physics) 115 (2010) A08101.
\newblock \href {http://dx.doi.org/10.1029/2009JA014848}
  {\path{doi:10.1029/2009JA014848}}.

\bibitem{2015SoPh..290..841D}
M.~{Dierckxsens}, K.~{Tziotziou}, S.~{Dalla}, I.~{Patsou}, M.~S. {Marsh}, N.~B.
  {Crosby}, O.~{Malandraki}, G.~{Tsiropoula}, {Relationship between Solar
  Energetic Particles and Properties of Flares and CMEs: Statistical Analysis
  of Solar Cycle 23 Events}, \solphys 290 (2015) 841--874.
\newblock \href {http://arxiv.org/abs/1410.6070} {\path{arXiv:1410.6070}},
  \href {http://dx.doi.org/10.1007/s11207-014-0641-4}
  {\path{doi:10.1007/s11207-014-0641-4}}.

\bibitem{2016JSWSC...6A..42P}
A.~{Papaioannou}, I.~{Sandberg}, A.~{Anastasiadis}, A.~{Kouloumvakos}, M.~K.
  {Georgoulis}, K.~{Tziotziou}, G.~{Tsiropoula}, P.~{Jiggens}, A.~{Hilgers},
  {Solar flares, coronal mass ejections and solar energetic particle event
  characteristics}, Journal of Space Weather and Space Climate 6~(27) (2016)
  A42.
\newblock \href {http://dx.doi.org/10.1051/swsc/2016035}
  {\path{doi:10.1051/swsc/2016035}}.

\bibitem{2016simi.conf...27M}
R.~{Miteva}, S.~W. {Samwel}, M.~V. {Costa-Duarte}, D.~{Danov}, {The online
  catalog of Wind/EPACT proton events}, in: K.~{Georgieva}, B.~{Kirov},
  D.~{Danov} (Eds.), Proceedings of the Eighth Workshop ''Solar Influences on
  the Magnetosphere, Ionosphere and Atmosphere'', 30 May-3 June 2016 Sunny
  Beach, Bulgaria. ISSN: 2367-7570, 2016, pp. 27--30.

\bibitem{2013JSWSC...3A..12V}
R.~{Vainio}, E.~{Valtonen}, B.~{Heber}, O.~E. {Malandraki}, A.~{Papaioannou},
  K.-L. {Klein}, A.~{Afanasiev}, N.~{Agueda}, H.~{Aurass}, M.~{Battarbee},
  S.~{Braune}, W.~{Dr{\"o}ge}, U.~{Ganse}, C.~{Hamadache}, D.~{Heynderickx},
  K.~{Huttunen-Heikinmaa}, J.~{Kiener}, P.~{Kilian}, A.~{Kopp},
  A.~{Kouloumvakos}, S.~{Maisala}, A.~{Mishev}, R.~{Miteva}, A.~{Nindos},
  T.~{Oittinen}, O.~{Raukunen}, E.~{Riihonen}, R.~{Rodr{\'{\i}}guez-Gas{\'e}n},
  O.~{Saloniemi}, B.~{Sanahuja}, R.~{Scherer}, F.~{Spanier}, V.~{Tatischeff},
  K.~{Tziotziou}, I.~G. {Usoskin}, N.~{Vilmer}, {The first SEPServer event
  catalogue \~{}68-MeV solar proton events observed at 1 AU in 1996-2010},
  Journal of Space Weather and Space Climate 3~(27) (2013) A12.
\newblock \href {http://dx.doi.org/10.1051/swsc/2013030}
  {\path{doi:10.1051/swsc/2013030}}.

\bibitem{SWE:SWE185}
A.~Posner, \href{http://dx.doi.org/10.1029/2006SW000268}{Up to 1-hour
  forecasting of radiation hazards from solar energetic ion events with
  relativistic electrons}, Space Weather 5~(5) (2007) n/a--n/a, s05001.
\newblock \href {http://dx.doi.org/10.1029/2006SW000268}
  {\path{doi:10.1029/2006SW000268}}.
\newline\urlprefix\url{http://dx.doi.org/10.1029/2006SW000268}

\bibitem{2009SpWea...7.4008L}
M.~{Laurenza}, E.~W. {Cliver}, J.~{Hewitt}, M.~{Storini}, A.~G. {Ling}, C.~C.
  {Balch}, M.~L. {Kaiser}, {A technique for short-term warning of solar
  energetic particle events based on flare location, flare size, and evidence
  of particle escape}, Space Weather 7 (2009) S04008.
\newblock \href {http://dx.doi.org/10.1029/2007SW000379}
  {\path{doi:10.1029/2007SW000379}}.

\bibitem{2013AdSpR..52.2102C}
R.~{Chandra}, N.~{Gopalswamy}, P.~{M{\"a}kel{\"a}}, H.~{Xie}, S.~{Yashiro},
  S.~{Akiyama}, W.~{Uddin}, A.~K. {Srivastava}, N.~C. {Joshi}, R.~{Jain}, A.~K.
  {Awasthi}, P.~K. {Manoharan}, K.~{Mahalakshmi}, V.~C. {Dwivedi}, D.~P.
  {Choudhary}, N.~V. {Nitta}, {Solar energetic particle events during the rise
  phases of solar cycles 23 and 24}, Advances in Space Research 52 (2013)
  2102--2111.
\newblock \href {http://dx.doi.org/10.1016/j.asr.2013.09.006}
  {\path{doi:10.1016/j.asr.2013.09.006}}.

\bibitem{2015ICRC....M}
R.~{Mewaldt}, C.~{Cohen}, G.~M. {Mason}, T.~{von Rosenvinge}, G.~{Li},
  C.~{Smith}, A.~{Vourlidas}, {Investigating the Causes of Solar-Cycle
  Variations in Solar Energetic Particle Fluences and Composition}, in: 34th
  ICRC The Hague, The Netherlands, 34th ICRC The Hague, The Netherlands, 2015.

\bibitem{2017SunGeo..12.0M}
R.~{Miteva}, S.~W. {Samwel}, M.~V. {Costa-Duarte}, O.~E. {Malandraki}, {Solar
  cycle dependence of Wind/EPACT protons, solar flares and coronal mass
  ejections}, Sun and Geosphere 12.

\bibitem{2013SSRv..175...53R}
D.~V. {Reames}, {The Two Sources of Solar Energetic Particles}, \ssr 175 (2013)
  53--92.
\newblock \href {http://arxiv.org/abs/1306.3608} {\path{arXiv:1306.3608}},
  \href {http://dx.doi.org/10.1007/s11214-013-9958-9}
  {\path{doi:10.1007/s11214-013-9958-9}}.

\bibitem{2016JASTP.146..171L}
B.~{Lavraud}, Y.~{Liu}, K.~{Segura}, J.~{He}, G.~{Qin}, M.~{Temmer}, J.-C.
  {Vial}, M.~{Xiong}, J.~A. {Davies}, A.~P. {Rouillard}, R.~{Pinto},
  F.~{Auch{\`e}re}, R.~A. {Harrison}, C.~{Eyles}, W.~{Gan}, P.~{Lamy},
  L.~{Xia}, J.~P. {Eastwood}, L.~{Kong}, J.~{Wang}, R.~F.
  {Wimmer-Schweingruber}, S.~{Zhang}, Q.~{Zong}, J.~{Soucek}, J.~{An},
  L.~{Prech}, A.~{Zhang}, P.~{Rochus}, V.~{Bothmer}, M.~{Janvier},
  M.~{Maksimovic}, C.~P. {Escoubet}, E.~K.~J. {Kilpua}, J.~{Tappin},
  R.~{Vainio}, S.~{Poedts}, M.~W. {Dunlop}, N.~{Savani}, N.~{Gopalswamy}, S.~D.
  {Bale}, G.~{Li}, T.~{Howard}, C.~{DeForest}, D.~{Webb}, N.~{Lugaz}, S.~A.
  {Fuselier}, K.~{Dalmasse}, J.~{Tallineau}, D.~{Vranken}, J.~G.
  {Fern{\'a}ndez}, {A small mission concept to the Sun-Earth Lagrangian L5
  point for innovative solar, heliospheric and space weather science}, Journal
  of Atmospheric and Solar-Terrestrial Physics 146 (2016) 171--185.
\newblock \href {http://dx.doi.org/10.1016/j.jastp.2016.06.004}
  {\path{doi:10.1016/j.jastp.2016.06.004}}.

\bibitem{2004JGRA..10912105G}
N.~{Gopalswamy}, S.~{Yashiro}, S.~{Krucker}, G.~{Stenborg}, R.~A. {Howard},
  {Intensity variation of large solar energetic particle events associated with
  coronal mass ejections}, Journal of Geophysical Research (Space Physics) 109
  (2004) A12105.
\newblock \href {http://dx.doi.org/10.1029/2004JA010602}
  {\path{doi:10.1029/2004JA010602}}.

\bibitem{2003ICRC....6.3305T}
A.~J. {Tylka}, C.~M.~S. {Cohen}, W.~F. {Dietrich}, S.~{Krucker}, R.~E.
  {McGuire}, R.~A. {Mewaldt}, C.~K. {Ng}, D.~V. {Reames}, G.~H. {Share},
  {Onsets and Release Times in Solar Particle Events}, International Cosmic Ray
  Conference 6 (2003) 3305.

\bibitem{1999ApJ...519..864K}
S.~{Krucker}, D.~E. {Larson}, R.~P. {Lin}, B.~J. {Thompson}, {On the Origin of
  Impulsive Electron Events Observed at 1 AU}, \apj 519 (1999) 864--875.
\newblock \href {http://dx.doi.org/10.1086/307415} {\path{doi:10.1086/307415}}.

\bibitem{2005A&A...442..673H}
K.~{Huttunen-Heikinmaa}, E.~{Valtonen}, T.~{Laitinen}, {Proton and helium
  release times in SEP events observed with SOHO/ERNE}, \aap 442 (2005)
  673--685.
\newblock \href {http://dx.doi.org/10.1051/0004-6361:20042620}
  {\path{doi:10.1051/0004-6361:20042620}}.

\bibitem{2014SoPh..289.2601M}
R.~{Miteva}, K.-L. {Klein}, I.~{Kienreich}, M.~{Temmer}, A.~{Veronig}, O.~E.
  {Malandraki}, {Solar Energetic Particles and Associated EIT Disturbances in
  Solar Cycle 23}, \solphys 289 (2014) 2601--2631.
\newblock \href {http://arxiv.org/abs/1402.1676} {\path{arXiv:1402.1676}},
  \href {http://dx.doi.org/10.1007/s11207-014-0499-5}
  {\path{doi:10.1007/s11207-014-0499-5}}.

\bibitem{2009EM&P..104..295G}
N.~{Gopalswamy}, S.~{Yashiro}, G.~{Michalek}, G.~{Stenborg}, A.~{Vourlidas},
  S.~{Freeland}, R.~{Howard}, {The SOHO/LASCO CME Catalog}, Earth Moon and
  Planets 104 (2009) 295--313.
\newblock \href {http://dx.doi.org/10.1007/s11038-008-9282-7}
  {\path{doi:10.1007/s11038-008-9282-7}}.

\bibitem{2009ApJ...691.1222R}
E.~{Robbrecht}, D.~{Berghmans}, R.~A.~M. {Van der Linden}, {Automated LASCO CME
  Catalog for Solar Cycle 23: Are CMEs Scale Invariant?}, \apj 691 (2009)
  1222--1234.
\newblock \href {http://arxiv.org/abs/0810.1252} {\path{arXiv:0810.1252}},
  \href {http://dx.doi.org/10.1088/0004-637X/691/2/1222}
  {\path{doi:10.1088/0004-637X/691/2/1222}}.

\bibitem{2015SoPh..290.1741R}
I.~G. {Richardson}, T.~T. {von Rosenvinge}, H.~V. {Cane}, {The Properties of
  Solar Energetic Particle Event-Associated Coronal Mass Ejections Reported in
  Different CME Catalogs}, \solphys 290 (2015) 1741--1759.
\newblock \href {http://arxiv.org/abs/1505.03071} {\path{arXiv:1505.03071}},
  \href {http://dx.doi.org/10.1007/s11207-015-0701-4}
  {\path{doi:10.1007/s11207-015-0701-4}}.

\bibitem{2013SoPh..288..241N}
N.~V. {Nitta}, M.~J. {Aschwanden}, P.~F. {Boerner}, S.~L. {Freeland}, J.~R.
  {Lemen}, J.-P. {Wuelser}, {Soft X-ray Fluxes of Major Flares Far Behind the
  Limb as Estimated Using STEREO EUV Images}, \solphys 288 (2013) 241--254.
\newblock \href {http://arxiv.org/abs/1304.4163} {\path{arXiv:1304.4163}},
  \href {http://dx.doi.org/10.1007/s11207-013-0307-7}
  {\path{doi:10.1007/s11207-013-0307-7}}.

\bibitem{2015SoPh..290.1947C}
I.~M. {Chertok}, A.~V. {Belov}, V.~V. {Grechnev}, {A Simple Way to Estimate the
  Soft X-ray Class of Far-Side Solar Flares Observed with STEREO/EUVI},
  \solphys 290 (2015) 1947--1961.
\newblock \href {http://arxiv.org/abs/1505.01649} {\path{arXiv:1505.01649}},
  \href {http://dx.doi.org/10.1007/s11207-015-0738-4}
  {\path{doi:10.1007/s11207-015-0738-4}}.

\bibitem{2008JGRA..113.1104H}
T.~A. {Howard}, D.~{Nandy}, A.~C. {Koepke}, {Kinematic properties of solar
  coronal mass ejections: Correction for projection effects in spacecraft
  coronagraph measurements}, Journal of Geophysical Research (Space Physics)
  113 (2008) A01104.
\newblock \href {http://dx.doi.org/10.1029/2007JA012500}
  {\path{doi:10.1029/2007JA012500}}.

\bibitem{2013ApJ...777..167D}
J.~A. {Davies}, C.~H. {Perry}, R.~M.~G.~M. {Trines}, R.~A. {Harrison},
  N.~{Lugaz}, C.~{M{\"o}stl}, Y.~D. {Liu}, K.~{Steed}, {Establishing a
  Stereoscopic Technique for Determining the Kinematic Properties of Solar Wind
  Transients based on a Generalized Self-similarly Expanding Circular
  Geometry}, \apj 777 (2013) 167.
\newblock \href {http://dx.doi.org/10.1088/0004-637X/777/2/167}
  {\path{doi:10.1088/0004-637X/777/2/167}}.

\bibitem{2003GeoRL..30lSEP3G}
N.~{Gopalswamy}, S.~{Yashiro}, A.~{Lara}, M.~L. {Kaiser}, B.~J. {Thompson},
  P.~T. {Gallagher}, R.~A. {Howard}, {Large solar energetic particle events of
  cycle 23: A global view}, \grl 30~(12) (2003) 8015.
\newblock \href {http://dx.doi.org/10.1029/2002GL016435}
  {\path{doi:10.1029/2002GL016435}}.

\bibitem{1982ApJ...261..710K}
S.~W. {Kahler}, {Radio burst characteristics of solar proton flares}, \apj 261
  (1982) 710--719.
\newblock \href {http://dx.doi.org/10.1086/160381} {\path{doi:10.1086/160381}}.

\bibitem{2001JGR...10620947K}
S.~W. {Kahler}, {The correlation between solar energetic particle peak
  intensities and speeds of coronal mass ejections: Effects of ambient particle
  intensities and energy spectra}, \jgr 106 (2001) 20947--20956.
\newblock \href {http://dx.doi.org/10.1029/2000JA002231}
  {\path{doi:10.1029/2000JA002231}}.

\bibitem{2013SoPh..282..579M}
R.~{Miteva}, K.-L. {Klein}, O.~{Malandraki}, G.~{Dorrian}, {Solar Energetic
  Particle Events in the 23rd Solar Cycle: Interplanetary Magnetic Field
  Configuration and Statistical Relationship with Flares and CMEs}, \solphys
  282 (2013) 579--613.
\newblock \href {http://dx.doi.org/10.1007/s11207-012-0195-2}
  {\path{doi:10.1007/s11207-012-0195-2}}.

\bibitem{2003psa..book.....W}
J.~V. {Wall}, C.~R. {Jenkins}, {Practical Statistics for Astronomers. Cambridge
  observing handbooks for research astronomers, vol. 3. Cambridge, UK:
  Cambridge University Press, 2003}, 2003.

\bibitem{2013CEAB...37..541M}
R.~{Miteva}, K.-L. {Klein}, S.~W. {Samwel}, A.~{Nindos}, A.~{Kouloumvakos},
  H.~{Reid}, {Radio Signatures of Solar Energetic Particles During the 23\^{}rd
  Solar Cycle}, Central European Astrophysical Bulletin 37 (2013) 541--553.
\newblock \href {http://arxiv.org/abs/1402.6442} {\path{arXiv:1402.6442}}.

\bibitem{2014EP&S...66..104G}
N.~{Gopalswamy}, H.~{Xie}, S.~{Akiyama}, P.~A. {M{\"a}kel{\"a}}, S.~{Yashiro},
  {Major solar eruptions and high-energy particle events during solar cycle
  24}, Earth, Planets, and Space 66 (2014) 104.
\newblock \href {http://arxiv.org/abs/1408.3617} {\path{arXiv:1408.3617}},
  \href {http://dx.doi.org/10.1186/1880-5981-66-104}
  {\path{doi:10.1186/1880-5981-66-104}}.

\end{thebibliography}

\end{document}